# Enhanced vector-based model for elastic bonds in solids

V.A. Kuzkin[1,2], A.M. Krivtsov[1,2]

kuzkinva@gmail.com, akrivtsov@bk.ru

[1]Peter the Great Saint Petersburg Polytechnic University, Polytechnicheskaya str. 29, St. Petersburg, 195251, Russia

[2] Institute for Problems in Mechanical Engineering RAS, Bolshoy pr. V.O. 61, St. Petersburg, 199178, Russia

Abstract

A model (further referred to as the enhanced vector-based model or EVM) for elastic bonds in solids, composed of bonded particles is presented. The model can be applied for a description of elastic deformation of rocks, ceramics, concrete, nanocomposites, aerogels and other materials with structural elements interacting via forces and torques. A material is represented as a set of particles (rigid bodies) connected by elastic bonds. Vectors rigidly connected with particles are used for description of particles orientations. Simple expression for potential energy of a bond is proposed. Corresponding forces and torques are calculated. Parameters of the potential are related to longitudinal, transverse (shear), bending, and torsional stiffnesses of the bond. It is shown that fitting parameters of the potential allows one to satisfy any values of stiffnesses. Therefore, the model is applicable to bonds with arbitrary length/thickness ratio. Bond stiffnesses are expressed in terms of geometrical and elastic properties of the bonds using three models: Bernoulli-Euler beam, Timoshenko beam, and short elastic cylinder. An approach for validation of numerical implementation of the model is presented. Validation is carried out by a comparison of numerical and analytical solutions of four test problems for a pair of bonded particles. Benchmark expressions for forces and torques in the case of pure tension/compression, shear, bending and torsion of a single bond are derived. This approach allows one to minimize the time required for a numerical implementation of the model.

## 1. Introduction

Discrete mechanical models are widely used for simulation of deformation and fracture of solids at different scale levels [1-15]. In the framework of discrete models, a solid is represented as a set of interacting particles. At nanoscale level, interactions between particles are usually described by interatomic potentials [1, 7]. At higher scale levels, interactions can be simulated by the so-called bonds connecting particles. The bonds either represent some additional glue-like material, connecting particles [8-12], or appear as a result of coarse-graining, e.g. of macromolecules [13,14], nanotube-based materials [15-16], etc. This model is widely used, for example, for simulations of granular solids, such as rocks, ceramics, concrete, nanocomposites, agglomerates, aerogels, etc.

Each bond causes forces and torques acting on the bonded particles. In the case of purely elastic bonds, forces and torques depend on mutual positions and orientations of the particles. In the two-dimensional case, construction of this dependence is relatively straightforward [3, 17], while in three-dimensional case it is a serious challenge. Several three-dimensional models for elastic bonds in solids are proposed in literature [9,15,16,20,21]. In paper [18], bonds are modeled by Timoshenko beams [19] connecting particles. The model has clear physical meaning, though its applicability is limited to bonds with relatively large length/thickness ratio. In paper [9], the so-called bonded particle model (BPM) is proposed. In the BPM, forces and torques are calculated using an incremental procedure. In this case, conservation of energy (required for simulation of elastic bonds) is not always guaranteed. In paper [20], an approach based on the decomposition of relative rotation of particles is proposed. Forces and torques are represented as functions of angles, describing relative rotation of the particles. Expression for the potential energy of the bond is not presented. Therefore, it is unclear if forces and torques [20] are conservative.

Another approach for simulation of elastic bonds (referred to as the vector-based model or V-model) is presented in paper [21]. In the V-model, potential energy of the bond is a function of the mutual position of the particles and vectors rigidly connected with particles. Expressions for forces and torques are derived from the potential energy. Therefore, the interactions caused by the bonds are truly conservative.

In the present paper, we further develop the approach proposed in paper [21]. Enhanced vector-based model (EVM) for elastic bonds is presented. A simple expression for the potential energy of the bond is proposed. Corresponding forces and torques are derived. Parameters of the potential energy are related to longitudinal, transverse (shear), bending, and torsional stiffnesses of the bond. The range of applicability of the EVM is discussed. An approach for validation of numerical implementation of the EVM is proposed.

## 2. Enhanced vector-based model (EVM) of an elastic bond

Consider a model of elastic bonds in a solid composed of bonded particles. In general, each particle can be bonded with any number of neighbors. The behavior of the bonds is assumed to be independent, i.e. pair interactions are considered. Therefore, for simplicity, two bonded particles $i$ and $j$ are considered below. Expressions for forces and torques acting on the particles $i$ and $j$ are derived.

Assume that the bond connects two points that belong to the particles. The points lie on the line connecting particles' centers in the initial (undeformed) state. Distances from these points to particles' centers of mass are denoted as $R_i$, $R_j$ respectively. For example, in the case shown in figure 1, the points lie on particles' surfaces and therefore $R_i$, $R_j$ are equal to particles' radii.

Orientations of the particles are described by orthogonal unit vectors $\mathbf{n}_{i1}$, $\mathbf{n}_{i2}$, $\mathbf{n}_{i3}$ and $\mathbf{n}_{j1}$, $\mathbf{n}_{j2}$, $\mathbf{n}_{j3}$, rigidly connected with particles $i$ and $j$ respectively. Here the first index corresponds to particle number, the second index corresponds to vector's number (see figure 1). In the undeformed state, the following relations are satisfied:

$$\mathbf{n}_{i1} = -\mathbf{n}_{j1} = \mathbf{e}_{ij}, \qquad \mathbf{n}_{i2} = \mathbf{n}_{j2}, \qquad \mathbf{n}_{i3} = \mathbf{n}_{j3}, \qquad (1)$$

where $\mathbf{e}_{ij} = \mathbf{r}_{ij}/r_{ij}$, $\mathbf{r}_{ij} = \mathbf{r}_j - \mathbf{r}_i$.

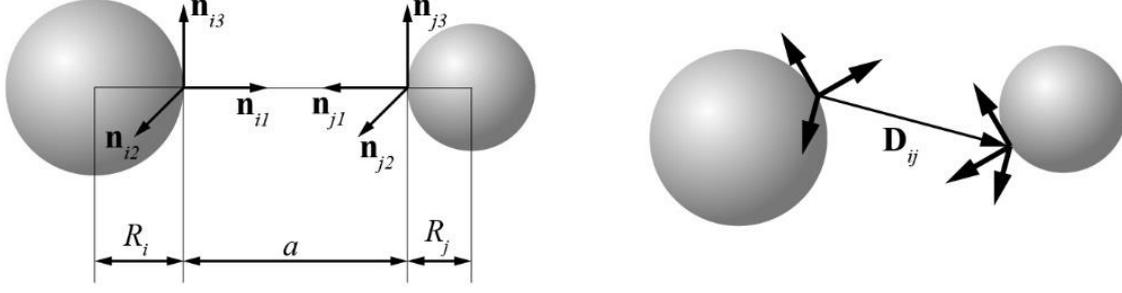

Figure 1: Two bonded particles in the undeformed state (left) and deformed state (right). Here and below *a* is an equilibrium length of the bond.

In general, potential energy of the bond is a function of vector $\mathbf{D}_{ij} \stackrel{\text{def}}{=} \mathbf{r}_{ij} + R_j \mathbf{n}_{j1} - R_i \mathbf{n}_{i1}$ and vectors $\mathbf{n}_{ik}$, $\mathbf{n}_{jm}$, $k, m = 1, 2, 3$. Vector $\mathbf{D}_{ij}$ is directed along the bond (see figure 1). It connects points with radius-vectors $\mathbf{r}_i + R_i \mathbf{n}_{i1}$, $\mathbf{r}_j + R_j \mathbf{n}_{j1}$. In the case $R_i = R_j = 0$, the bond connects particle centers ($\mathbf{D}_{ij} = \mathbf{r}_{ij}$).

The expression for potential energy of the bond, $U$, proposed in paper [21] can be simplified. In the framework of the EVM, the following expression for potential energy is used:

$$U = \frac{B_1}{2}(D_{ij} - a)^2 + \frac{B_2}{2}(\mathbf{n}_{j1} - \mathbf{n}_{i1}) \cdot \mathbf{d}_{ij} + B_3 \mathbf{n}_{i1} \cdot \mathbf{n}_{j1} - \frac{B_4}{2}(\mathbf{n}_{i2} \cdot \mathbf{n}_{j2} + \mathbf{n}_{i3} \cdot \mathbf{n}_{j3}), \quad (2)$$

where $D_{ij} = |\mathbf{D}_{ij}|$, $\mathbf{d}_{ij} = \mathbf{D}_{ij}/D_{ij}$; $B_1, B_2, B_3, B_4$ are parameters of the model related to stiffnesses of the bond (see formula (11)).

Formula (2) is significantly simpler than corresponding expression in the V-model [21]. The main difference between these expressions is in the last term. In formula (2), the last term contributes to both bending and torsion of the bond. In paper [21] the potential energy is designed in such a way that at small deformations bending and torsion are described by two independent terms. This "decomposition" leads to an unnecessarily complicated expression for potential energy.

Forces $\mathbf{F}_{ij}$ and torques $\mathbf{M}_{ij}$, $\mathbf{M}_{ji}$ acting between the particles are calculated as follows [6,13,21]:

$$\mathbf{F}_{ij} = -\mathbf{F}_{ji} = \frac{\partial U}{\partial \mathbf{r}_{ij}}, \qquad \mathbf{M}_{ij} = \sum_{k=1}^{3} \frac{\partial U}{\partial \mathbf{n}_{ik}} \times \mathbf{n}_{ik}, \qquad \mathbf{M}_{ji} = \sum_{k=1}^{3} \frac{\partial U}{\partial \mathbf{n}_{jk}} \times \mathbf{n}_{jk}.$$

Here torque $\mathbf{M}_{ij}$ is calculated with respect to center of mass of particle *i*. Taking the derivatives yields:

$$\mathbf{F}_{ij} = B_1(D_{ij} - a)\mathbf{d}_{ij} + \frac{B_2}{2D_{ij}}(\mathbf{n}_{j1} - \mathbf{n}_{i1} - \mathbf{d}_{ij} \cdot (\mathbf{n}_{j1} - \mathbf{n}_{i1})\mathbf{d}_{ij}),$$
$$\mathbf{M}_{ij} = R_i\mathbf{n}_{i1} \times \mathbf{F}_{ij} - \frac{B_2}{2}\mathbf{d}_{ij} \times \mathbf{n}_{i1} + \mathbf{M}^{TB},$$
$$\mathbf{M}_{ji} = R_j\mathbf{n}_{j1} \times \mathbf{F}_{ji} + \frac{B_2}{2}\mathbf{d}_{ij} \times \mathbf{n}_{j1} - \mathbf{M}^{TB}, \qquad (3)$$
$$\mathbf{M}^{TB} = B_3\mathbf{n}_{j1} \times \mathbf{n}_{i1} - \frac{B_4}{2}(\mathbf{n}_{j2} \times \mathbf{n}_{i2} + \mathbf{n}_{j3} \times \mathbf{n}_{i3}).$$

If the bond connects particles' centers, expressions (3) take the form:

$$\mathbf{F}_{ij} = B_1(r_{ij} - a)\mathbf{r}_{ij} + \frac{B_2}{2r_{ij}}(\mathbf{n}_{j1} - \mathbf{n}_{i1} - \mathbf{e}_{ij} \cdot (\mathbf{n}_{j1} - \mathbf{n}_{i1})\mathbf{e}_{ij}),$$
$$\mathbf{M}_{ij} = -\frac{B_2}{2}\mathbf{e}_{ij} \times \mathbf{n}_{i1} + \mathbf{M}^{TB}, \qquad (4)$$
$$\mathbf{M}_{ji} = \frac{B_2}{2}\mathbf{e}_{ij} \times \mathbf{n}_{j1} - \mathbf{M}^{TB},$$

where $\mathbf{e}_{ij} = \mathbf{r}_{ij} / r_{ij}$, $\mathbf{M}^{TB}$ is defined by equation (3).

**Remark.** In contrast to the BPM [9], forces and torques (3), (4) are defined by instantaneous positions and orientations of the particles. Therefore in the framework of the EVM, more accurate symplectic methods (methods conserving energy) for numerical integration of equations of motion can be used.

## 3. Benchmarks: tension, shear, bending, and torsion of the bond

The behavior of solids composed of bonded particles interacting via forces and torques (3), (4), in general, is quite complicated. Therefore, validation of numerical implementation of the EVM is not straightforward. In the present section, a simple approach for validation is proposed. Forces and torques acting on two bonded particles are calculated in the case of tension/compression, shear, bending, and torsion of the bond (see figure 2). The resulting expressions can be used as benchmarks for a validation of computer implementation of the model. Additionally, the relations between parameters of the model and stiffnesses of the bond are derived. For simplicity it is assumed that the bond connects particle centers ($R_i = R_j = 0$).

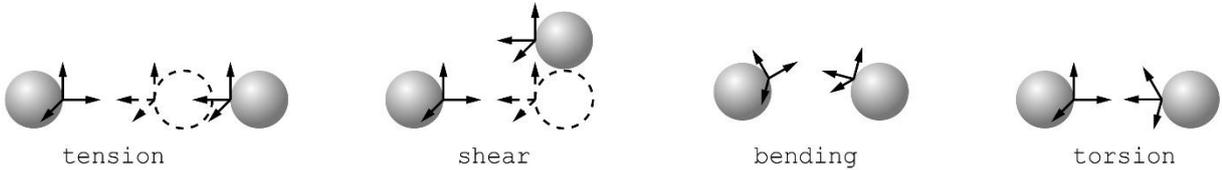

Figure 2: Deformations of the bond and corresponding orientation of vectors, connected with the particles.

### A. Tension and shear of the bond

Consider pure tension of the bond connecting particles $i$ and $j$. Particles are displaced along vector $\mathbf{r}_{ij}$. In this case, forces and torques (4) take the form

$$\mathbf{F}_{ij} = B_1(r_{ij} - a)\mathbf{e}_{ij}, \qquad \mathbf{M}_{ij} = \mathbf{M}_{ji} = 0. \qquad (5)$$

One can see that in the case of pure tension the behavior of the bond is identical to the behavior of a linear spring with stiffness $B_1$. Therefore longitudinal stiffness of the bond $c_A = B_1$.

Consider pure shear of the bond. Assume that position of particle $i$ is fixed and particle $j$ has displacement $u\mathbf{k}$, where the unit vector $\mathbf{k}$ is orthogonal to the initial direction of the bond $\mathbf{e}_0$. In this case, forces and torques (4) acting between the particles are the following:

$$\mathbf{F}_{ij} = B_1(r_{ij} - a)\mathbf{e}_{ij} - \frac{B_2}{r_{ij}}(\mathbf{e}_0 - \mathbf{e}_0 \cdot \mathbf{e}_{ij}\mathbf{e}_{ij}),$$
$$\mathbf{M}_{ij} = \mathbf{M}_{ji} = -\frac{B_2}{2}\mathbf{e}_{ij} \times \mathbf{e}_0. \quad (6)$$

Rewriting formulae (6) using the geometrical relations

$$\mathbf{e}_0 \cdot \mathbf{e}_{ij} = \frac{a}{r_{ij}}, \quad \mathbf{e}_{ij} \cdot \mathbf{k} = \frac{u}{r_{ij}}, \quad \mathbf{e}_{ij} \times \mathbf{e}_0 = \frac{u}{r_{ij}}\mathbf{k} \times \mathbf{e}_0,$$
$$r_{ij} = \sqrt{a^2 + u^2}, \quad (7)$$

yields

$$\mathbf{F}_{ij} \cdot \mathbf{e}_0 = B_1 a\left(1 - \frac{a}{\sqrt{a^2 + u^2}}\right) - \frac{B_2 u^2}{(a^2 + u^2)^{\frac{3}{2}}},$$
$$\mathbf{F}_{ij} \cdot \mathbf{k} = B_1 u\left(1 - \frac{a}{\sqrt{a^2 + u^2}}\right) + \frac{B_2 u a}{(a^2 + u^2)^{\frac{3}{2}}}, \quad (8)$$
$$\mathbf{M}_{ij} = \mathbf{M}_{ji} = -\frac{B_2 u}{2\sqrt{a^2 + u^2}}\mathbf{k} \times \mathbf{e}_0.$$

Linearization of the second formula from (8) with respect to $u$ yields $\mathbf{F}_{ij} \cdot \mathbf{k} \approx c_D u$, where $c_D = B_2/a^2$ is the shear stiffness of the bond.

Thus, parameters of the model $B_1$ and $B_2$ are proportional to longitudinal, $c_A$, and shear, $c_D$, stiffnesses of the bond respectively.

### B. Bending and torsion of the bond

Consider pure bending of the bond. Assume that positions of particles $i$ and $j$ are fixed. Particles $i, j$ are rotated in opposite directions by angle $\varphi$ around vector $\mathbf{n}_{i2} = \mathbf{n}_{j2}$. In this case, the force vanishes and torques are equal with opposite sign:

$$\mathbf{M}_{ij} = -\mathbf{M}_{ji} = -\left(\frac{B_2}{2}\sin\varphi + B_3 \sin 2\varphi + \frac{B_4}{2}\sin 2\varphi\right)\mathbf{n}_{i2},$$
$$\mathbf{F}_{ij} = 0. \quad (9)$$

In the case of small rotations of the particles $\mathbf{M}_{ij} = -\mathbf{M}_{ji} \approx -2c_B \varphi \mathbf{n}_{i2}$, where $c_B = B_2/4 + B_3 + B_4/2$ is the bending stiffness of the bond.

Consider torsion of the bond. Assume that position and orientation of particle $i$ is fixed and particle $j$ is rotated around $\mathbf{r}_{ij}$ by angle $\varphi$. Then

$$\mathbf{M}_{ij} = -\frac{B_4}{2}(\mathbf{n}_{j2} \times \mathbf{n}_{i2} + \mathbf{n}_{j3} \times \mathbf{n}_{i3}) = -B_4 \sin\varphi \, \mathbf{e}_{ij},$$
$$\mathbf{M}_{ij} = -\mathbf{M}_{ji}, \quad \mathbf{F}_{ij} = 0. \quad (10)$$

It is seen that torsional stiffness, $c_T$, is equal to parameter $B_4$.

Formulas (5), (8), (9), (10) can be used for validation of numerical implementation of the EVM.

## 4. Calibration of the model

According to formulas (5), (8), (9), (10), stiffnesses of the bond are related to parameters $B_k$ of the potential energy (2) by the following simple formulas:

$$c_A = B_1, \qquad c_D = \frac{B_2}{a^2}, \qquad c_B = \frac{B_2}{4} + B_3 + \frac{B_4}{2}, \qquad c_T = B_4. \tag{11}$$

It is seen that *any* values of longitudinal, shear, bending, and torsional stiffnesses of the bond can be fitted by a proper choice of parameters $B_k$. Therefore the EVM is applicable to bonds with an arbitrary length/thickness ratio. Thus in spite of the simplification, it has the same advantage as the V-model [21].

In the present section, parameters of the model, $B_k$, are expressed in terms of geometrical parameters of bonds and mechanical properties of bonding material.

### 4.1. Long bonds

Mechanical behavior of relatively long bonds can be described by the beam theory [19]. We derive the relation between parameters of the model, $B_k$, and characteristics of massless Timoshenko beam connecting particles. Assume that the beam has equilibrium length $a$, constant cross-section, and isotropic bending stiffness. Longitudinal, shear, bending, and torsional stiffnesses of Timoshenko beam are derived in paper [21]:

$$c_A = \frac{ES}{a}, \qquad c_D = \frac{12kEJS}{a(kSa^2 + 24J(1+\nu))},$$
$$c_B = \frac{EJ}{a}, \qquad c_T = \frac{GJ_p}{a}, \tag{12}$$

where $E$, $G$, $\nu$ are Young's modulus, shear modulus, and Poisson's ratio of the bonding material; $S$, $J$, $J_p$ are area, moment of inertia, and polar moment of inertia of the cross section; $k$ is a dimensionless shear coefficient [19]. Then formulas (11) and (12) yield the relation between parameters of the model and characteristics of Timoshenko beam:

$$B_1 = \frac{ES}{a}, \qquad B_2 = \frac{12kaEJS}{kSa^2 + 24J(1+\nu)}, \qquad B_4 = \frac{GJ_p}{a},$$
$$B_3 = \frac{EJ}{a} - \frac{B_2}{4} - \frac{B_4}{2}. \tag{13}$$

In the limit $k \to \infty$, Timoshenko beam is equivalent to Bernoulli-Euler beam. Corresponding relation between the parameters has the form:

$$B_1 = \frac{ES}{a}, \qquad B_2 = \frac{12EJ}{a}, \qquad B_3 = -\frac{2EJ}{a} - \frac{GJ_p}{2a},$$
$$B_4 = \frac{GJ_p}{a}. \tag{14}$$

If parameters $B_k$ are determined by formula (14), then under small deformations the model is equivalent to the Bernoulli-Euler beam connecting particles.

Thus formulas (13), (14) can be used for calibration of the EVM, provided that bonds are relatively long, i.e. Bernoulli-Euler or Timoshenko beam models are applicable.

**4.2. Short bonds**

Beam models described above are inapplicable if longitudinal and transverse stiffnesses are of the same order ($c_A/c_D \sim 1$). For example, in materials composed of glued particles, bonds are usually short [8-12] and the stiffnesses are comparable. Similarly, for covalent bonds in graphene $c_A/c_D \sim 2$ [22]. Therefore in the present section an alternative model [21] is used for calibration of parameters $B_k$.

The bond is approximated by a short cylinder with equilibrium length $a$. Material of the cylinder is linearly elastic and isotropic. Then longitudinal, shear, bending, and torsional stiffnesses of a short bond are related to characteristics of the bonding material as follows (see paper [21]):

$$c_A = \frac{(1-\nu)}{(1+\nu)(1-2\nu)}\frac{ES}{a}, \qquad c_D = \frac{GS}{a},$$
$$c_B = \frac{(1-\nu)}{(1+\nu)(1-2\nu)}\frac{EJ}{a}, \qquad c_T = \frac{GJ_p}{a}. \tag{15}$$

Then formulas (11) yield expressions, connecting parameters of the model with characteristics of the bond:

$$B_1 = \frac{(1-\nu)}{(1+\nu)(1-2\nu)}\frac{ES}{a}, \qquad B_2 = GSa, \qquad B_4 = \frac{GJ_p}{a},$$
$$B_3 = \frac{(1-\nu)}{(1+\nu)(1-2\nu)}\frac{EJ}{a} - \frac{B_2}{4} - \frac{B_4}{2}. \tag{16}$$

Thus in the case of short bonds, calibration of the model is carried out using formulas (16).

**5. Conclusions**

In the present paper, enhanced vector-based model (EVM) for simulation of elastic bonds in solids has been presented. The potential energy of the bond was represented via vectors, rigidly connected with bonded particles. Forces and torques caused by the bond were calculated. Relations between parameters of the model and bond stiffnesses were derived. It was shown that fitting parameters of the model allows one to satisfy *any* values of longitudinal, shear, bending, and torsional stiffnesses. Therefore the model is applicable to bonds with an *arbitrary* length/thickness ratio.

Validation of numerical implementation of the model has been discussed. It has been proposed to simulate four types of deformations (tension, shear, bending, and torsion) in a two particle system and to calculate corresponding forces and torques. Then numerical results can be compared with simple analytical expressions derived in the present paper. This approach allows one to minimize the time required for a numerical implementation of the model.

Finally, we note that the EVM can be used at different length scales. At nano scale level, covalent bonds (e.g. in graphene [4-6,22] and molybdenum disulfide [23]) can be simulated. At micro scale level, the EVM can be incorporated into coarse-grained models of macromolecules [13,14], nanotubes [15,16], aerogels, ceramics [11] etc. At macro scale level, the EVM can be applied to geomechanical problems, e.g. simulation of hydraulic fracturing in naturally fractured reservoirs [24].

**Acknowledgements**

This work was supported by Ministry of Education and Science of the Russian Federation within the framework of the Federal Program "Research and development in priority areas for the development of the scientific and technological complex of Russia for 2014-2020" (activity 1.2), grant No. 14.575.21.0146 of September 26, 2017, unique identifier: RFMEFI57517X0146.